\documentclass[epsfig,11pt]{article}
\usepackage{epsfig,citesort}

      \input{amssym.def}
      \input{amssym}
      \newcommand {\mm}[1] {\ifmmode{#1}\else{\mbox{\(#1\)}}\fi}
      \newcommand{\real} {\mm{{\Bbb R}}}

\newcommand{\utwi}[1]{\mbox{\boldmath $ #1$}}
\newcommand{\ba}{{\utwi{a}}}

\newcommand{\bc}{{\utwi{c}}}

\newcommand{\bs}{{\utwi{s}}}

\newcommand{\bw}{{\utwi{w}}}

\begin{document}
      \title{\bf Empirical Potential Function for Simplified Protein
      Models: Combining Contact and Local Sequence-Structure
      Descriptors}
\author{Jinfeng Zhang$^1$, Rong Chen$^{1,2,3}$, and Jie Liang$^{1}$\\
\thanks{Corresponding author.  Phone: (312)355--1789, fax: (312)996--5921, email:
{\tt jliang@uic.edu}
}\\
$^1$Department of Bioengineering,\\
$^2$Department of Information \& Decision Science, \\ University of Illinois at
Chicago, Chicago,IL, USA\\
$^3$Department of Business Statistics \& Econometrics, \\
Peking University,
Beijing, P.R. China\\
}

\date{\today, Accepted by {\it Proteins}}
\maketitle
\abstract{An effective potential function is critical for protein
structure prediction and folding simulation.  Simplified protein
models such as those requiring only $C_\alpha$ or backbone atoms are
attractive because they enable efficient search of the conformational
space.  We show residue specific reduced discrete state models can
represent the backbone conformations of proteins with small RMSD
values.  However, no potential functions exist that are designed for
such simplified protein models.  In this study, we develop optimal
potential functions by combining contact interaction descriptors and
local sequence-structure descriptors.  The form of the potential
function is a weighted linear sum of all descriptors, and the optimal
weight coefficients are obtained through optimization using both
native and decoy structures.  The performance of the potential
function in test of discriminating native protein structures
from decoys is evaluated using several benchmark decoy sets. Our
potential function requiring only backbone atoms or $C_\alpha$ atoms
have comparable or better performance than several residue-based
potential functions that require additional coordinates of side chain
centers or coordinates of all side chain atoms.  By reducing the
residue alphabets down to size 5 for local structure-sequence
relationship, the performance of the potential function can be further
improved.  Our results also suggest that local sequence-structure
correlation may play important role in reducing the entropic cost of
protein folding.

}

\vspace*{.8in}

\noindent
{\bf Keywords: } Decoy discrimination; discrete state model; potential
function; protein structure prediction; simplified protein models;
local sequence-structure relationship.
\vspace*{.2in}

\newpage

\section{Introduction}
\label{sec:intro}

Protein folding is a fundamental problem in molecular biology
\cite{Anfinsen73_S,Dill90,Dobson03_N}. The thermodynamic hypothesis of
protein folding postulates that the native state of a protein has
 lowest free energy under physiological conditions. Under this
hypothesis, protein structure prediction, folding simulation, and
protein design all depend on the use of a potential function.  In
protein structure prediction, the potential function is used either to
guide the conformational search process, or to select
a structure from a set of possible sampled candidate structures.

There are several challenging difficulties in computational studies of
protein structures.  The search space of protein conformation is
enormous, and the native structure cannot be identified by exhaustive
enumeration.  This is the well-known ``Levinthal's paradox''
\cite{Levinthal68}.  In addition, we do not yet have full
understanding of all the physical factors and how they work
collectively in folding proteins and maintaining protein stability.
Simplified protein models provide an attractive approach that helps to
overcome these two difficulties
\cite{HeadGordonBrown03_COSB,KolinskiSkolnick04_Polymer}.  Based on
simplified protein representation, these models can effectively reduce
the complexity in conformational search.  They are also valuable for
isolating and identifying the most relevant factors contributing to
protein folding, without the need to model an overwhelming amount of
detailed atomistic information required when all-atom representation
of protein structure is used.

There are several key technical issues in using simplified protein
models.  First, which form of the simplified protein representation
would contain the needed relevant information? Second, what
descriptors should we choose to extract the necessary information?
Finally, how do we construct a potential function using these
descriptors so near native structures will have lower energy than
others?
In this study, we develop an empirical potential function for
simplified protein models at the residue-level.  Our work fills an
important gap.  Existing empirical potential functions require either
all-atom representation of protein structures
\cite{LuSkolnick01_P,ZhouZhou02_PS,McConkeySobolev03_PNAS,BastollaVendruscolo01_P},
or the coordinates of the geometric center of side chains
\cite{MiyazawaJernigan96_JMB,TobiElber00_P}, which require explicit
model of side chain atoms.  Currently, there is no accurate potential
function designed for simplified protein models requiring only
$C_\alpha$ or backbone atoms.  An effective potential function is
essential for efficient conformational search, for evaluation of
sampled structures, and for realization of the capabilities of a
well-designed simplified protein model.

In this study, we choose the discrete state off-lattice model
originally developed by Park and Levitt as the reduced
representation for protein structures \cite{ParkLevitt95_JMB}.
The states of this model for each residue is parameterized by a
bond angle and a torsion angle.  This model has been shown to work
well in modeling protein structures, at the same time maintaining
a low complexity \cite{ParkLevitt95_JMB}.  We extend the original
model and develop  a set of optimal discrete states for each amino
acids through clustering of the observed angles in native protein
structures.

For these simplified off-lattice discrete state models, we follow a
novel approach to develop descriptors.  We use both two body residue
contact interactions and the local sequence-structure information of
two sequence nearest neighboring residues.  Contact interactions
capture several basic physical forces important for protein folding,
including hydrophobic interactions, hydrogen bonding, charge
interactions, and disulfide bonding interactions
\cite{Cline02_Proteins}.  Contact interactions have been used in many
empirical potentials
\cite{MiyazawaJernigan96_JMB,BastollaVendruscolo01_P,TobiElber00_P,LuSkolnick01_P,ZhouZhou02_PS,LiHuLiang03_P,McConkeySobolev03_PNAS}.
The local sequence-structure correlation of residues capture the
propensity of small sequences adopting specific local spatial
structures.  The existence of such propensity has been well recognized
and it has been used in protein structure prediction
\cite{SimonsBaker97_JMB,HunterSubram03_P}, in remote homology
detection \cite{HouBystroff04_P}, and in discriminating native
structures from decoys
\cite{Shortle02_PS,KolodnyLevitt02_JMB,PeiGrishin04_P,LezonMaritan04_P}.
The non-overlapping nature of these two types of descriptors indicates
that they contain different information.  To our best knowledge,
potential function developed in this study is the first to combine
both types of descriptors.

There are two approaches for developing an empirical potential.  One
approach uses only native protein structures and apply statistical
analysis to extract information important for protein stability
\cite{MiyazawaJernigan96_JMB,LuSkolnick01_P,ZhouZhou02_PS,McConkeySobolev03_PNAS}.
The other approach uses both native protein structures and decoy
conformations and apply optimization (or machine learning) techniques
to derive a potential function that separates native structures from
decoy structures
\cite{LazaridisKarplus00_COSB,HaoScheraga99_COSB,BucheteStraubThirumalai04_COSB,HuLiLiang04_Bioinfo}.
The approach based on statistical analysis has the drawback of
assuming explicitly or implicitly an unrealistic reference state such
as random mixture model \cite{MiyazawaJernigan96_JMB}, and ignoring
chain connectivity \cite{ThomasDill96_JMB}.  The approach based on
optimization involves deriving parameters from a set of training
proteins and decoys, and is attractive because it incorporates
information contained in the decoy structures that are absent in
native structures.  The collection of a very large number of decoy
structures plays the role of the reference state in the statistical
methods.
In addition, the optimization approach allows
more flexibility in combining descriptors of different physical
nature.  To develop potential function by optimization, it is
important to select a small or moderate number of descriptors to avoid
over-fitting the training examples. For this purpose, we
systematically develop several reduced alphabet of amino acid residues
for both contact interactions and for local sequence-structure
descriptors.

The potential function we developed here are tested for discrimination
of native protein structures from several benchmark sets of decoy
non-protein conformations.  For all the decoys tested, the performance
of our potential is comparable or better than several well-known
residue-level potential functions that requiring more complex protein
representations.  Our paper is organized as following: first, we
introduce the simplified representation of protein structures.
Second, we discuss the reduction of amino acid alphabet for
neighboring interaction patterns. We then describe the descriptors and
the form of the potential function, along with the optimization method
to derive the weight vector of the potential function. This is
followed by description of the performance of the potential function
in discriminating native structures from decoys.  Finally, we conclude
this paper with discussion.

\vspace*{-.1in}

\section{Model and Methods}
\label{sec:meth}

\subsection{Representation of protein structures.}
\begin{figure}[tbh]
\centerline{\epsfig{figure=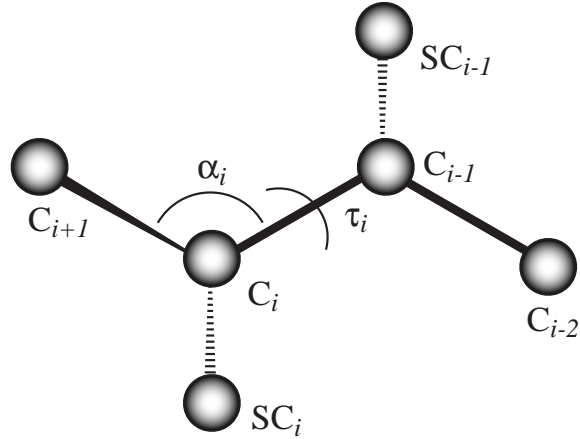,width=3.0in}}
\caption[Discrete state model]{\sf Discrete state model. $\alpha_i$
and $\tau_i$ angles are shown for residue $i$. Bond angle
$\alpha_i$ at position $i$ is formed by $C_{i-1}$, $C_{i}$, and
$C_{i+1}$. Torsion angle $\tau_i$ is the dihedral angle of the two
planes formed by atoms ($C_{i-2}$,$C_{i-1}$,$C_i$) and ($C_{i-1}$,
$C_i$, $C_{i+1}$).} \label{Fig:model}
\end{figure}

{\sc \textbf{Discrete state model.}}  We use an off-lattice discrete
state model to represent the protein structure
\cite{ParkLevitt95_JMB}. In addition to $C_\alpha$ atoms, we use one
additional atom $SC$ to model all side chain atoms, which is attached
to the main chain $C_\alpha$ atoms, as shown in
Figure~\ref{Fig:model}. The distance between adjacent $C_\alpha$ atoms
is fixed to 3.8 \AA. The distances between $C_\alpha$ and side chain
atoms, as well as the radius of each side chain atom depend on the
residue type.  Their values are taken from reference
\cite{ParkLevitt96_JMB}. There is no additional increase in the degree
of freedom due to the introduction of the side chain atom.

Similar to the fact that the position of $C_\beta$ atom of the side
chain in a protein is uniquely determined from the positions of
backbone atoms as the atoms connected to carbons have fixed angles,
the position of the residue-dependent side chain atom $SC$ in this
model is determined from backbone $C_\alpha$ atom
\cite{ParkLevitt96_JMB}.

There are 20 different types of atoms altogether (one $C_\alpha$ for
backbone and glycine, and 19 different $SC$ atoms for different side
chains). The backbone structure of a protein can be described by the
bond angles $\alpha_i$ and torsion angles $\tau_i$ at the $i$-th
$C_\alpha$ position (Figure~\ref{Fig:model}). The overall three
dimensional structure is completely determined by the set of angles
\{($\alpha_i$, $\tau_i$)\} at each $C_\alpha$ position, except the
terminal residues.

\begin{figure}[tbh]
\centerline{\epsfig{figure=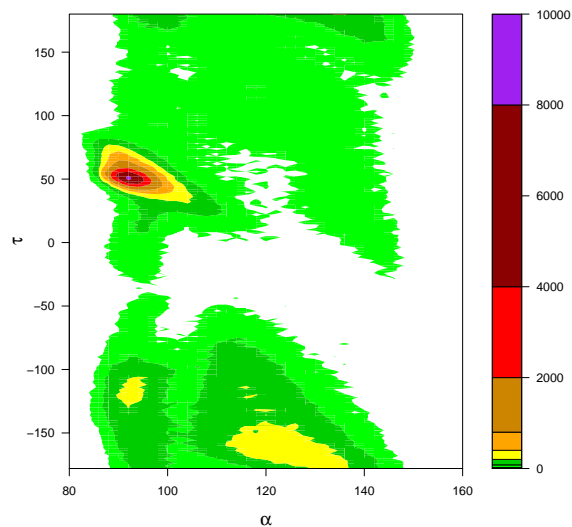,width=3in}}
\caption[Distributions of bond angles and dihedral
angles]{\sf Distribution of $\alpha$ and $\tau$ angles taken by 20 natural amino acids in native proteins. } \label{Fig:aa_dist}
\end{figure}

{\sc \textbf{$\alpha$ and $\tau$ angles.}}  To find the desirable
number of states and the associated ($\alpha$, $\tau$) values of amino
acid residues for the discrete state model, we obtained the
distribution of $\alpha$ and $\tau$ angles in 1,318 non-homologous
X-ray protein structures from CulledPDB \cite{CulledPdb}
(Figure~\ref{Fig:aa_dist}), where the sequence identity between any
pairs of proteins is less than 30 percent, and the resolution of the
structures is better than 2 \AA. \ Analogous to the Ramachandran plot,
the distribution of $\alpha$ and $\tau$ angles also has densely and
sparsely populated regions, which correspond to different secondary
structure types The distribution of ($\alpha$, $\tau$) angles differs
for different amino acids.

{\sc \textbf{Reduced discrete states.}}
For each residue, we obtain a Cartesian coordinate system by taking
the plane formed by $C_{i-2}$, $C_{i-1}$, and $C_i$ as the $x-y$
plane, and placing the origin at $C_i$. The vector from $C_{i-1}$ to
$C_i$ is taken as the direction of $x$-axis.  After normalizing the
bond length between $C_i$ and $C_{i+1}$ to unit length, we cluster the
positions of $C_{i+1}$ atoms for all residues of the same type, which are
taken as $C_i$.
$k$-mean clustering is applied to group points on the unit sphere into
$k$ (from 3 to 10) clusters for each amino acid residue type, where $k$
corresponds to the number of states for the amino acids. The centers
of the clusters are then measured for the $\alpha$ and $\tau$ angles,
which are taken as the optimized values of the discrete states of the
amino acids. The values of discrete state angles for the 4 state model
are listed in Table~\ref{tab:dis_angle}.  The values for $k$-state
model of $k = 5 - 10$ are listed in supplementary material.

\renewcommand{\arraystretch}{.7}
\begin{table}
\begin{center}
\caption[Discrete state angles]
{\sf Values of discrete state angles for 4 state model.
}
\label{tab:dis_angle}
\vspace*{.1in}
\begin{tabular}{|r|r|r|r|r|r|r|r|r|}
\hline
A.A. & ${\alpha}_1$ & $\tau_1$ & $\alpha_2$ & $\tau_2$ & $\alpha_3$ & $\tau_3$ &
$\alpha_4$ & $\tau_4$ \\
\hline
A & 104.9 & -112.3 & 91.80 & 52.10 & 125.3 & -175.7 & 134.8 & 86.03 \\
C & 112.4 & -107.0 & 98.07 & 45.22 & 123.9 & -170.3 & 120.3 & 111.4 \\
D & 106.3 & -108.9 & 96.31 & 45.00 & 113.1 & -168.9 & 113.7 & 107.1 \\
E & 106.9 & -106.3 & 94.64 & 49.22 & 117.9 & -165.8 & 116.3 & 113.7 \\
F & 112.2 & -105.9 & 98.50 & 46.15 & 122.9 & -166.7 & 120.9 & 116.0 \\
G & 108.2 & -96.99 & 102.3 & 36.01 & 124.9 & -165.0 & 133.1 & 110.4 \\
H & 108.3 & -101.6 & 98.73 & 45.03 & 119.6 & -164.7 & 122.9 & 112.8 \\
I & 110.3 & -108.5 & 95.57 & 47.44 & 119.2 & -163.9 & 116.4 & 115.4 \\
K & 108.1 & -108.9 & 95.28 & 48.98 & 116.9 & -164.8 & 117.0 & 115.8 \\
L & 110.3 & -110.7 & 94.31 & 48.84 & 117.7 & -163.9 & 115.3 & 114.6 \\
M & 110.8 & -107.1 & 94.50 & 49.24 & 121.7 & -166.0 & 118.8 & 116.6 \\
N & 106.2 & -109.6 & 96.27 & 41.75 & 116.9 & -172.9 & 122.3 & 99.00 \\
P & 110.1 & -104.3 & 93.65 & 41.43 & 105.0 & -163.4 & 100.0 & 131.7 \\
Q & 108.4 & -109.7 & 94.70 & 49.15 & 119.3 & -167.2 & 117.8 & 112.1 \\
R & 108.4 & -112.9 & 93.20 & 49.67 & 121.4 & -174.3 & 127.6 & 93.57 \\
S & 114.5 & -103.3 & 99.38 & 49.30 & 120.8 & -163.8 & 119.0 & 122.1 \\
T & 115.5 & -105.7 & 99.57 & 47.03 & 121.6 & -165.2 & 121.2 & 122.0 \\
V & 111.1 & -110.3 & 96.96 & 46.87 & 121.0 & -165.1 & 117.5 & 116.4 \\
W & 112.4 & -105.3 & 96.64 & 48.12 & 121.5 & -166.4 & 117.7 & 119.9 \\
Y & 113.3 & -103.5 & 99.41 & 45.47 & 124.2 & -166.5 & 119.5 & 118.3 \\
\hline
\end{tabular}
\end{center}
\end{table}

\begin{figure}[tbh]
\centerline{\epsfig{figure=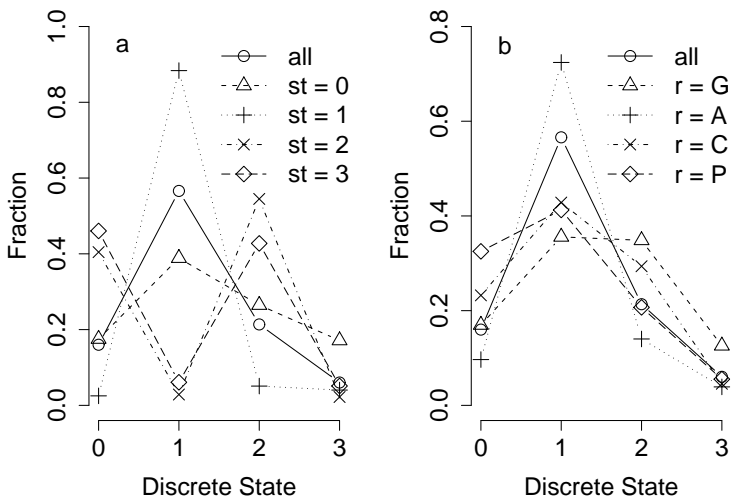,width=4in}}
\caption[Neighboring Residue Effect]{\sf The distribution of
discrete states of alanine calculated from 1,318 non-homologous
x-ray protein structures given (a) the discrete state or (b) the
type of preceding residue. Y-axis shows the fraction of different
discrete state of alanine as labeled on X-axis. Labels in legend:
{\sf all}, marginal distribution of discrete state of alanine
regardless the state or residue type of the preceding residue;
{\sf st}, the discrete state of the preceding residue; {\sf r},
the type of the preceding residue. We applied a bootstrap
procedure using 1,000 sampling with replacement following
\cite{Adamian03_JMB} to obtain the confidence intervals for the
data shown. The $95\%$ confidence intervals for the above data are
all within $(f-0.03,f+0.03)$, where $f$ is the fraction shown on
the figure.} \label{Fig:nbrEff}
\end{figure}

To study the effect of the preceding residue to the distribution of
the discrete state of each type of amino acids, we plotted the
conditional distribution of the discrete state for each amino acid
residue given the state of the preceding residue. The results for
alanine is shown in Figure~\ref{Fig:nbrEff}a, which shows that the
distribution of the discrete state of alanine is affected
significantly by the state of the preceding residue.  This
distribution is also affected by the type of preceding residue, as
shown in Figure~\ref{Fig:nbrEff}b. Similar effects are observed in
all other residues.

\begin{figure}[t]
\centerline{\epsfig{figure=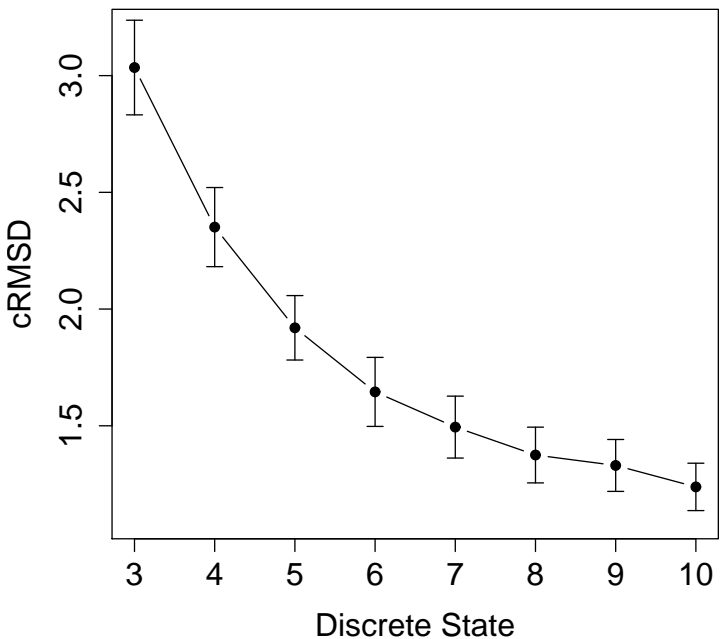,width=4in}}
\caption[Average RMSD of NNS]{\sf Average RMSD values of the best 2,000
discrete state models {\it vs.} the number of states. The average and standard deviation of RMSD
values are calculated from a set of 70 proteins obtained from \cite{Fain02_PS}.  The average length of these proteins are 137.}
\label{Fig:rmsd_ns_best}
\end{figure}

{\sc \textbf{Mapping of X-ray structures to discrete state models.}}
The conformational space associated with a discrete state
representation is different from the continuous conformational space
of a protein structure in $\real^3$. To represent a protein in the
simplified discrete space, we need to map a protein structure from the
continuous conformational space to a structure in a discrete space,
with the requirement that it must be the one most similar to the real
protein structure among all possible structures in the simplified
conformational space by some similarity measure. In this study, we use
both {\it global structural similarity\/} and {\it local structural
similarity\/} criteria. To generate globally similar discrete
structures to an X-ray structures, we use a heuristic ``build-up''
algorithm first introduced by Park and Levitt
\cite{ParkLevitt95_JMB}. In this method, the protein structure is
constructed in single residue increments starting from the
N-terminus. At each step of construction, only a fixed number of $m$
structures with the lowest RMSD from the partial X-ray structure are
retained. When a residue is added to the growing chain, all $k$
possible states on each of the retained chains are examined for
conformation similarity to X-ray structure. This gives $k \times m$
possible conformations at each step for a $k$-state model, of which
the best $m$ conformations are retained for the next step of
construction. The representatives obtained from the build-up method
are the ones among the final $m$ full protein candidate structures
that has the lowest global RMSD values from the native structure.
With this method, we obtained $m=5,000$ discrete structures for each
proteins in the set of 70 representative proteins obtained in
\cite{Fain02_PS}, with average length of 136 residues for 3 to 10
discrete states. The average RMSD values of the best fitted structures
for each discrete state are shown in Figure~\ref{Fig:rmsd_ns_best}. We
also fitted 978 proteins with less than 500 residues taken from the
1,318 proteins in the CulledPDB with $m = 2,000$ conformations for
each proteins for state 3 to 6 and obtained similar results with
slightly larger average RMSD values. In general, the average RMSD
values of the best models to the native structures is about 2.3 \AA\
for 4-state model, 1.9 \AA \ for 5-state model, 1.6 \AA \ for 6-state
model, and near 1.0 \AA\ for 10-state model.

The high quality of the discrete state models when fitted to X-ray
structures indicates that a model with four to six states is
sufficient to generate near native structures with low RMSD values,
\emph{i.e.}, $<3$ \AA\, to native structures. In the rest of the
paper, we use 4-state model for its simplicity.

To generate discrete state model by local structural similarity, each
residue is simply assigned a discrete state that is most similar to
its local ($\alpha$,$\tau$) angle in X-ray structure.  The resulting
structure has maximum local similarity to X-ray structures
\cite{LezonMaritan04_P}.

\subsection{Descriptors.}

{\sc \textbf{Simplified amino acid alphabet $\Sigma_1$ by neighbor
interactions.}}  Early protein synthesis has been thought to involve a
reduced amino acid alphabet
\cite{RiddleBaker97_NSB}. Previous work has shown that a small $\beta$-sheet
protein, the SH3 domain, can be encoded by a reduced 5-letter amino
acid alphabet \cite{RiddleBaker97_NSB}.  Despite the dramatic changes
in sequence, the folding rates of the protein encoded by the reduced
alphabet are very close to that of the naturally occurring SH3 domain
\cite{RiddleBaker97_NSB}. Various reduced amino acid alphabets have
been obtained previously based on the analysis of amino acid
substitution matrix, contact propensity, or information theory
\cite{RiddleBaker97_NSB,LiHuLiang03_P,LiWang03_PE,CannataValle02_Bioinfo,LevyMurphy00_PE,WangWang99_NSB}.
The resulting reduced amino acid alphabets are useful in protein
folding studies and in identifying consensus sequences from multiple
alignment \cite{Dill85_Biochem,SagotSoldano97_TCS}. However, there
has been no attempt to derive reduced alphabet based on the local
sequence-structure relationship of the amino acids. By recognizing
strong similarities of different amino acids in their local spatial
interaction patterns, one can summarize the relationship between local
sequence and structure more succinctly and accurately.  In addition,
using reduced alphabet also alleviates problems arising from the use
of a limited data set of protein structures and decoys to derive
empirical potential functions. Such a simplified alphabet would also
be useful in representation of protein structures, in building
fragment libraries, in prediction of local structures from local
sequences, and in generating protein like conformations using chain growth
method
\cite{SchlickGanTropsha00_JCP,LiangZhangChen02_JCP,ZhangLiu02_JCP,ZhangCTL03_JCP,ZhangChenChenLiang04_JCP}.

Different amino acids have different distributions of $\alpha$ and
$\tau$ angles. However, amino acids that are similar in geometrical
shapes or chemical properties often share similar patterns in
the distribution of $\alpha$ and $\tau$ angles.
The posterior distribution of discrete state angles for a residue
given the preceding residue's type and discrete state provides
characteristic information of local structure of residues. The observed
neighboring residue effect shown in Figure~\ref{Fig:nbrEff}
indicates that the type and geometry
represented by discrete state of one residue also affect the geometry
of its adjacent residues. These observations prompt us to simplify the
twenty amino acids alphabet to a smaller alphabet.
To derive the
simplified alphabet, we estimate the first order state transition
probability of residues in native protein structures as described below.

\begin{figure}[tb]
\centerline{\epsfig{figure=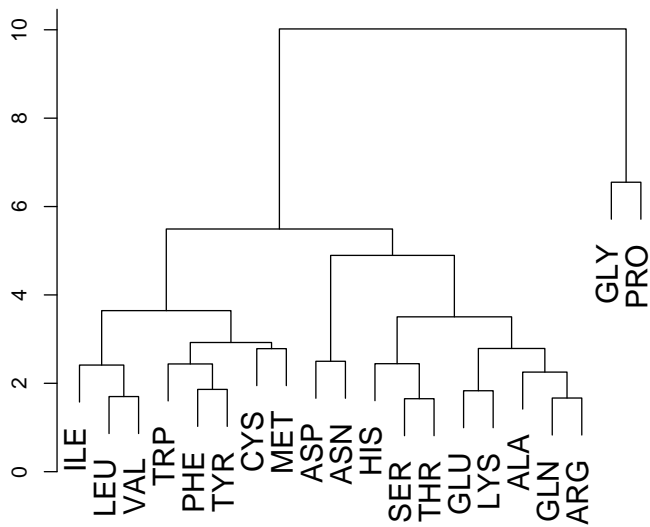,width=4.0in}}
\caption[Hierarchical clustering of amino acids using their
neighboring residue interaction patterns]{\sf Hierarchical clustering
of amino acids using their neighboring residue interaction
patterns. Glycine and proline are residues having highest tendency
of being at turn and loop positions, which are separated from the
rest of amino acids. The rest 18 amino acids are clustered into
two groups roughly according to their hydrophobicities.  }
\label{Fig:aa_clus}
\end{figure}

A protein structure can be represented uniquely by a sequence of
$(a,x)$, where $a$ is amino acid residue type and $x$ is the discrete
state. For a four-state model with 20 amino acid types, the total
number of possible descriptors for one residue position is $20 \times
4 = 80$. For simplicity, we use $s \in [1 \ldots 80]$ to represent the state
a residue may take, {\it i.e.}, the discrete conformational state and
the amino acid type. We define the first order state transition
probability $p_{s_1,s_2}$ as: $p_{s_1,s_2} = p[s_2|s_1] =
p[(a_{2},x_{2})|(a_1,x_1)]$ and
calculate the transition matrix from 1,318
non-homologous proteins. We then
cluster different residue
types based on the transition probabilities. Each type of residue
corresponds to a vector of 320 transition probabilities (moving
from one of the four states associated with this particular type of
residue to one of the 80 different residue type and state combinations).
We define the distance between two amino acid types as
the Euclidean distance between the corresponding
transition probabilities vectors. Results of clustering of amino acids using
this distance metric are shown in Figure~\ref{Fig:aa_clus}. The twenty
amino acids can be divided into two distinct groups, with one group
containing only glycine and proline. Our clustering results are very
different from those using other criteria.  Geometrically, it is
intuitive that glycine and proline have no side chains and are
different from other residues.
For this study, we further group the residues into an alphabet
$\Sigma_1$ of 5 letters with $\mathcal{A}=\{A,E,K,Q,R,S,H,T\}$, $
\mathcal{B}=\{C,I,V,L,M,W,F,Y\}$, $\mathcal{C}= \{D,N\}$, $
\mathcal{D}=\{G\}$, and $\mathcal{E}=\{P\}$.

{\sc \textbf{Simplified amino acid alphabet $\Sigma_2$ for contact
propensity: incorporating additional descriptors.}} We use a
different reduced alphabet for contact interactions. In this work, we
take an alphabet $\Sigma_2$ based on results published in
\cite{LiHuLiang03_P}, where an alphabet of ten residue types are
chosen as following: $\{I,L,V\}, \{C\}, \{A\}, \{G\}, \{N,Q,S,T\},
\{P,H\}, \{M,F\}$, $\{W,Y\}$, $\{D,E\}$, $\{K,R\}$. This reduced
alphabet is used for simplification of contact descriptors.

{\bf Descriptor set ${\cal D}_1$. } To encode the information
contained in local-sequence and local-structure of sequence
neighboring residues, we use the two discrete state taken by two
consecutive residues, $(x_{i-1},x_i,a_{i-1},a_i)$ as descriptors.  The
number of possible descriptor values for a pair of residues is $(4
\times 4) \times (5 \times 5) = 400$, since there are 4 discrete
conformational states and 5 simplified amino acid types in alphabet
$\Sigma_1$.

{\bf Descriptor set ${\cal D}_2$. } Contact interactions in a
protein structure can be uniquely defined once the contact criterion
is given. With 20 types of atoms (19 side chain atoms and 1 backbone
atom), the number of different types of contacts, or contact
descriptors is $210$. This set of descriptors is denoted as ${\cal
D}_{21}$.  When using 10 reduced atom types derived from the
simplified amino acid alphabet $\Sigma_2$ with 10 amino acid types,
the number of contact descriptors is $55$, and we denoted it as ${\cal
D}_{2,2}$.  Because of the reduction in the number of descriptors, we
can afford to incorporate additional descriptors that are more
informative. As an exploratory study, we further distinguish each
pairwise contact type by the sequence separation $d_{i,j}$ of the two
contacting residues, where $d_{i,j}=|j-i|$. We group $d_{i,j}$ values into
three bins, with bin 1 for $d_{i,j}=4$, bin 2 for $d_{i,j}=5$, and bin
3 for $d_{i,j} >5$.  Thus, the total number of contact descriptors
becomes $55 \times 3 =165$. This set is denoted as ${\cal D}_{2,3}$.

{\sc \textbf{Combining contact and local sequence-structure
descriptors.  }}  We have experimented with three different sets of
descriptors obtained by combining 400 local sequence-structure
descriptors ${\cal D}_1$ with the three different sets of contact
descriptors ${\cal D}_{2,1}, {\cal D}_{2,2} $ and ${\cal D}_{2,3}$.
CALS (Contact And Local Sequence-structure) is the set of 610
descriptors combining ${\cal D}_1$ and ${\cal D}_{2,1}$.  RCALS1
(Reduced Contact And Local Sequence-structure 1) is the set of 455
descriptors combining ${\cal D}_1$ and ${\cal D}_{2,2}$.  RCALS2
(Reduced Contact And Local Sequence-structure 2) is the set of 565
descriptors combining ${\cal D}_1$ and ${\cal D}_{2,3}$.

We also study the property of using contact descriptor ${\cal
D}_1$ only (denoted as C potential) and local sequence-structure descriptors
${\cal D}_{21}$ only (denoted as LS potential).

{\sc \textbf{Calculation of the contact descriptors from a
structure.}}  If all backbone atoms are present in a given structure,
we obtain each $SC$ atom position  by extending the bond
length between $C_\alpha$ and $C_\beta$ atoms to a residue dependent value
along a fixed direction as in
\cite{ParkLevitt96_JMB}, and the new position of the $C_\beta$ is
taken as the position of side chain atom $SC$. If only $C_\alpha$
atoms is present, we estimate the $SC$ position for side chain atom
following the approach of \cite{ParkLevitt96_JMB}, where the
coordinates of side chain atom at position $i$ is approximately
determined by the coordinates of $C_\alpha$s at position $i-1$, $i$,
and $i+1$. After all $C_\alpha$ and $C_\beta$ atoms have been placed,
we calculate the contact descriptors by simply measuring the pair-wise
distance of atoms.  In our calculations, explicit information from
side chain atoms of a PDB structure is never used.

To derive local sequence-structure descriptors, we transform the
structure to a discrete state model using local fit as described
earlier. The local sequence-structure descriptors are calculated
directly from the discrete representation.

\subsection{Empirical potential function.}

Potential functions based on physical model (such as {\sc Charmm} and
{\sc Amber} \cite{CHARMM,AMBER}) require all-atom representations of
protein structures to model detailed physical forces and therefore are
inappropriate for simplified protein representations.  With a proper
representations of protein sequence and structure, and a set of
descriptors specified, we have a description function, $\bc =
f(\bs,\ba)$, which takes a pair of structure $\bs$ and sequence $\ba$
$(\bs,\ba)$, and maps it to a descriptor vector $\bc$. The next step
is to decide on the form of the potential function $E = H(\bc)$, which
maps the vector $\bc$ to a real valued energy or score, $E$.

The form of a potential function in this study is a linear combination
of the descriptors: $ H(\bc) = \bw \cdot \bc,$ {\it i.e.}, the inner
product of the descriptor vector $\bc$ and the weight vector $\bw$.
The energy landscape of an empirical potential function defined for
simplified protein model is inevitably different from the true energy
landscape of a real protein. For tasks such as protein structure
prediction, the minimum requirement is that the structures in the
conformational space of simplified protein model that are closest to
the native structure have the globally minimum energy
values.
Developing such a potential function is challenging, as it is
not even known that whether near native conformations in simplified
protein representation can be the most stable conformations in the
full conformational space under any particular potential functions
\cite{Betancourt03_P}.  Despite this uncertainty, there is still a
great deal of interest and work in developing optimized potential
functions that stabilizes native proteins in simplified protein models
\cite{TobiElber00_Proteins_1,ChiuGoldstein98_FD,MirnyShkh96_JMB,HuLiLiang04_Bioinfo,Vendruscolo98_JCP,Dima00_PS}.
This work is a continuation of efforts in this direction.

We obtain weight vector $\bw$ using optimization method
\cite{TobiElber00_Proteins_1,ChiuGoldstein98_FD,MirnyShkh96_JMB,HuLiLiang04_Bioinfo,Vendruscolo98_JCP,Dima00_PS}.
For
our linear potential functions, the basic requirement is: $ \bw \cdot
(\bc_N - \bc_D) + b < 0 $, where $\bc_N$ and $\bc_D$ are the native
descriptor vector and the decoy descriptor vector for one protein, and
$b \geq 0 $ is the energy gap between a native and decoy
structure that should exists. Each pair of native vector and decoy vector serves as one
inequality constraint. All of the constraints jointly define a convex
polyhedron $P$ for feasible weight vectors $\bw$'s. If $P$ is not
empty, there could be an infinite number of choices of $\bw$, all with
perfect discrimination \cite{HuLiLiang04_Bioinfo}. To find a weight
vector $\bw$ that is optimal, one can choose the weight vector $\bw$
that minimizes the variance of score gaps between decoys and natives
\cite{TobiElber00_Proteins_1}, or minimizing the $Z$-score of the
native protein and an ensemble of decoys
\cite{ChiuGoldstein98_FD,MirnyShkh96_JMB}, or maximizing the ratio
$R$ between the width of the distribution of the score and the
average score difference between the native state and the unfolded
ones \cite{Goldstein92_PNAS}. Previous works using perception
learning and other optimization techniques
\cite{FriedrichsWolynes89_Science,Goldstein92_PNAS,TobiElber00_Proteins_1,Vendruscolo98_JCP,Dima00_PS}
showed that often effective linear sum potential functions can be
obtained.

Here we obtain the optimal weight vector $\bw$ by solving the
following primal quadratic programming problem:
\begin{eqnarray}
\mbox{Minimize } & \frac{1}{2} || \bw||^2
\\
\mbox{subject to} & \bw \cdot (\bc_N - \bc_D) + b < 0 \mbox{ for
all } N \in {\mathcal N} \mbox{ and } D \in {\mathcal D}.
\label{Eqn:PrimalLinear}
\end{eqnarray}
The solution maximizes the distance $b/||\bw||$ of the plane
$(\bw, b)$ to the origin \cite{ScholkopfSmola02}. We use a support
vector machines (SVM)
for this task
\cite{SVM_Light}.

{\bf Potential function studied.} Based on the five different sets of
descriptors described above, we study the following five different
potential functions: CALSP (potential function based on the CALS
descriptor set), RCALSP1 (potential function based on the RCALS 1
descriptor set), RCALSP2 (potential function based on the RCALS 2
descriptor set), CP (potential function based on the C descriptor
set), and LSP (potential function based on the LS descriptor set).

\subsection{Data set for discrimination test.}

{\sc \textbf{Proteins database.}} We select 978 non-homologous
proteins from CulledPdb \cite{CulledPdb}, with the criteria that the
sequence identity is less than 30\%, the resolution of X-ray
structures is smaller than 2 \AA, and the R factor is smaller than
0.25. In addition, using a compactness parameter $z_\alpha$ developed
in \cite{ZhangCTL03_JCP}, we require that all have $z_\alpha$ values
greater than 3.0, so the compactness of the protein is that of the
single domain globular proteins. This compactness constraint excludes
proteins with extended conformations.  These proteins are unlikely to
be stable on their own, and usually requires protein-protein
interactions or protein-DNA interactions
\cite{HuLiLiang04_Bioinfo,BastollaVendruscolo01_P}.

{\sc \textbf{Gapless threading decoys.}} We use gapless threading to
generate a total of about 60 millions of decoys
\cite{MaiorovCrippen92_JMB}.  A three fold cross validation is applied
to train the potential function and test its performance.

{\sc \textbf{Decoys generated by Loose {\em et. al.} (LKF decoy
set)}.} This set of decoys are generated by Loose, Klepeis, and
Floudas using the program of DYANA, which takes as input the sequence
of a protein, along with information about its secondary structure
that gives bounds for the distances and torsion angles between atoms
\cite{LooseFloudas04_P,DYANA}. DYANA minimizes the energy of the
structure and then simulates a sharp increase in temperature, with a
step using molecular dynamics simulation that allows the shape of the
protein to change. The protein is then slowly cooled down, or
annealed, and its energy is again minimized to give the output
structure. Decoys for 185 proteins were downloaded from the authors'
website.  About 200 decoy structures for each protein are available to
us \cite{LooseFloudas04_P}.

{\sc \textbf{Decoys generated by Baker {\em et. al.} (Baker decoy
set).}} This set of decoys has 41 proteins. All decoys are generated
by the program Rosetta \cite{Rosetta99_P}. Several different protocols
are combined to produce the decoy set, which has the following
properties: (1) It contains conformations for a wide variety of
different proteins; (2) it contains conformations close ($<4$ \AA) to
the native structure; (3) it consists of conformations that are at
least near local minima for a reasonable potential function, so they
cannot be trivially excluded based on obviously non protein like
features; and (4) it is produced by a relatively unbiased procedure
that does not use information from the native structure during
conformational search \cite{TsaiBaker03_P}.

{\sc \textbf{4State\_reduced set.}} This decoy test set contains
native and near-native conformations of seven sequences, along
with about 650 misfolded structures for each sequence. Park and
Levitt generated the positions of $C_\alpha$ in these decoys by
exhaustively enumerating 10 selectively chosen residues in each
protein using a 4-state off-lattice model. All other residues
were assigned the $\phi$/$\psi$ values based on the best fit of a
4-state model to the native chain. Conformations in the decoy sets
all have low scores by a variety of potential functions, and low
root-mean square distance (RMSDs) to the native structures
\cite{SamudralaLevitt00_PS}.

{\sc \textbf{Lattice\_ssfit set.}} The Lattice\_ssfit set contains
conformations for eight small proteins generated by {\em ab
initio} protein structure prediction methods. The conformational
space of a sequence was exhaustively enumerated on a tetrahedral
lattice. A lattice-based potential function was used to select the
10,000 best-scoring conformations. Park and Levitt fitted
secondary structures to these conformations using a 4-state model.
The 10,000 conformations were further scored with a combination of
an all-atom potential function, a hydrophobic compactness
function, and a one-point per residue potential function. The
2,000 best-scoring conformations for each protein were selected as
decoys for this data set
\cite{SamudralaLevitt99_PSB,XiaLevitt00_JCP}.

{\sc \textbf{LMDS set.}} The local minima decoy set (LMDS)
contains decoys derived from the experimentally obtained secondary
structures of 10 small proteins belonging to diverse structural
classes. Each decoy is a local minimum of a ``hand-made'' energy
function. The authors generated ten thousand initial conformations
for each protein by randomizing the torsion angles of the loop
region \cite{Fletcher70_CJ}. The adjacent local minima were found
by truncated Newton-Raphson minimization in torsion space. Each
protein is represented in the decoy set by its 500 lowest energy
local minima.

\section{Results}
\subsection{Performance on gapless threading decoys.}
The performance of the potential function on decoys generated by
gapless threading is listed in Table~\ref{tab:glt_pfm}. A three-fold
cross validation is employed to test the potential function, where all
of the 978 proteins are randomly divided into three groups, and two
groups and their associated decoys are used in turn for training and
one group for testing. Among the 978 proteins, CALSP (see above) has
only 6 proteins misclassified, which corresponds to an accuracy of
99\%.
A protein is misclassified if there is one or more decoy structure(s) for that
protein with a lower score than that of the native structure.
 The potential functions RCALSP1 and RCALSP2 also give good
performance, with only 5 and 3 proteins misclassified, respectively.
We also tested potential functions containing only contact (CP) and
only local sequence-structure (LSP) descriptors.  Clearly, combining
both type of information in CALSP, RCALSP1, and RCALSP2 is much better
than using CP or LSP alone. By comparing the performance of CP and
LSP, we can also see that contact descriptors are more informative in
discriminating native structures from decoys than local
sequence-structure descriptors.

We also compare our potential functions with several other residue
based potential functions, including those  developed by Tobi {\em
et.al.} (TE13) \cite{TobiElber00_P}, Miyazawa \& Jernigan (MJ)
\cite{MiyazawaJernigan96_JMB}, and Bastolla {\em et.\ al}.\ (BV)
\cite{BastollaVendruscolo01_P}. Although these potential functions
are residue based potential functions, they need all-atom
representation since they either need to calculate the side chain
geometric centers or need to compute explicit atom-atom contacts.
CALSP, RCALSP1, and RCALSP2 are the only potential functions, that
can be applied on representations with only $C_\alpha$ and
$C_\beta$ atoms. Since $C_\beta$ position is completely determined
by coordinates of backbone atoms, these potential functions also
work for representation with only backbone atoms.
 The results for other potential functions are obtained from
tables in \cite{HuLiLiang04_Bioinfo}, where the authors followed
the original literature of contact definition, cut-off values, as
well as using the original potential parameters.  The training
data and test data in \cite{HuLiLiang04_Bioinfo} were obtained
from the {\sc Whatif} database \cite{WHATIF}, while the set of 978
proteins are obtained from the {\sc cullPDB} dataset.  Although
direct comparisons using exactly the same set of proteins is
impossible, the results listed in Table~\ref{tab:glt_pfm} indicate
that despite using a much simplified representation, CALSP has
comparable or better performance than other residue level
potential functions requiring more detailed representations.

\begin{table}[tb] % one column table
\caption[Performance on gapless threading decoys]{\sf Performance of
residue based potential functions in decoy discrimination. The
number of mis-classifications are listed. CALSP: Contact And Local
Sequence-structure Potential; RCALSP1: Reduced Contact And Local
Sequence-structure Potential with 455 descriptors; RCALSP2:
Reduced Contact And Local Sequence-structure Potential
incorporating contact order information with 565 descriptors; CP:
Contact potential using only contact component of CALSP; LSP:
Local sequence-structure potential using only local
sequence-structure component of CALSP; TE13: potential function
developed by Tobi \& Elber \cite{TobiElber00_P}; BV: Potential function developed by
Bastolla \& Vendruscolo \cite{BastollaVendruscolo01_P}; MJ: Potential function
developed by
Miyazawa \& Jernigan \cite{MiyazawaJernigan96_JMB}. AB: Computation of potential
function needs
only $C_\alpha$ and $C_\beta$ atoms; SCC: Computation of potential
function needs side chain center; AA: Computation of potential
function needs all-atom representation.}

\label{tab:glt_pfm} \vspace*{.1in}
\begin{center}
 \begin{tabular}{|l|r|r|}
\hline Potential Function  & Complexity     & Mis-classified Proteins \\
\hline CALSP       & AB             & 6/978          \\
\hline RCALSP1     & AB             & 5/978          \\
\hline RCALSP2     & AB             & 3/978          \\
\hline CP          & AB             & 24/978         \\
\hline LSP         & AB             & 249/978        \\ \hline
\hline TE13        & SCC            & 7/194          \\
\hline BV          & AA             & 2/194          \\
\hline MJ          & SCC            & 85/194         \\
\hline
\end{tabular}
\end{center}
\end{table}

\subsection{Performance on other decoy sets.}
{\bf LKF Set.}
When the potential function obtained from training using gapless
threading decoys is tested on other decoy sets, the performance of
discrimination is rather poor. This is not surprising, since it is
well known that gapless threading decoys are less challenging than
explicit decoys generated by different energy minimization
protocols. Potential functions derived by optimization frequently use
more realistic decoys. We therefore develop a new version of  potential function
CALSP based on
training with explicitly generated decoy conformations.  The LKF
and Baker decoy sets are used since these are the only ones with
relatively large
number of proteins and decoys (185 protein, 36,840 decoys for LKF
decoy set, and 41 proteins, 76,224 decoys for Baker decoy set,
respectively).

We use a four fold cross validation for the LKF decoy set, where all
of the 185 proteins are randomly divided into four groups, and three
groups are used in turn for training and one group for testing. No
gapless threading decoys are included in training.  As a comparison,
the performance of the original LKF potential on 151 of the 185
proteins are listed in \cite{LooseFloudas04_P}. For these 151
proteins, 140 proteins collected from test sets of different cross
validations are ranked as number 1 by our CALSP potential function
with an average z-score of 6.42, and 137 proteins ranked as number 1
by RCALSP1 with an average $z$-score of 6.15. As a comparison,
potential function LKF has 93 protein ranked as number 1 as reported
in \cite{LooseFloudas04_P}, with an average z-score of 3.08. Potential
function TE13 has 64 protein ranked as number 1 as reported in
\cite{LooseFloudas04_P}, with an average z-score of 2.01
(Table~\ref{tab:lkf_baker}).

Because the Baker decoy set contains only 41 proteins, which is too
small for a 4-fold cross validation test, we carried out leave-one-out
tests, again without including any gapless threading decoys during
training. Even though only 40 proteins are available for training each
time, our results are encouraging: we have 28 proteins ranked number
1, with an average z-score of 4.16 (Table~\ref{tab:lkf_baker}).

\begin{table}[tb] % one column table
\caption[Performance of CALSP on LKF and Baker decoys]{\sf
The number of mis-classifications using CALSP and other
residue based potential functions. CALSP: Contact And Local
Sequence-structure Potential; RCALSP1: Reduced Contact And Local
Sequence-structure Potential with 455 descriptors; TE13: potential
function developed by Tobi \& Elber; LKF: potential function developed
by Loose, Klepeis, and Floudas. \\ $*$ $z$-score is defined as $(\bar E
- E_n)/ \sigma$, where $\bar E$ is the average score of the decoys for
a protein, $E_n$ is the score of native conformation, and $\sigma$ is
the standard deviation of the scores of decoys.  \\ $**$ Result
obtained from \cite{LooseFloudas04_P}. It is not trained on LKF decoy
set.}

\label{tab:lkf_baker} \vspace*{.1in}
\begin{center}
 \begin{tabular}{|l|r|r|}
\hline Potential Function & Mis-classified Proteins & $z$-score$^*$\\
\hline \multicolumn{3}{|l|}{\bf LKF decoy set}     \\
\hline CALSP                & 11/151  & 6.42       \\
\hline RCALSP1               & 14/151  & 6.15       \\
\hline LKF                  & 58/151  & 3.08       \\
\hline TE13$^{**}$             & 87/151  & 2.01       \\
\hline \multicolumn{3}{|l|}{\bf Baker decoy set}   \\
\hline CALSP & 13/41 & 4.16         \\
\hline
\end{tabular}
\end{center}
\end{table}

For both LKF decoy sets and the Baker decoy set, we found that
inclusion of gapless threading decoys does not offer significant
improvement in performance. As discussed in
\cite{HuLiLiang04_Bioinfo}, this is because only a small number of
training examples will contribute as to determine the boundary between
proteins and decoys. In the study of LKF and Baker decoys, few such
training examples come from gapless threading decoys when the combined
training sets are used. This confirms earlier observation that decoys
from gapless threading are indeed less challenging. High quality
decoys are very much in need for the development of potential
functions by optimization methods.

{\bf Other Decoy Sets: 4State-reduced, LMDS, and Lattice-ssfit.}
We also test the CALSP potential function using the {\sc
4State-reduced} decoy set, {\sc LMDS} decoy set, and {\sc
Lattice\_ssfit} decoy set. Because the number of proteins in these
decoy sets are relatively small, we combined several training
sets, including gapless threading decoys, the near native
structures produced by the greedy build-up method and decoys from
LKF decoy set. Performance of CALSP on these decoy sets is listed
in Table~\ref{tab:dec_pfm}. We compare CALSP with three other
residue-based potential functions, namely, TE13, LL
\cite{LiHuLiang03_P}, and MJ. Performance of CALSP in general is
better or comparable to other potential functions. It performs
better than other potential functions on LMDS decoy set. Again,
although all potential functions are of residue level, only CALSP can
be applied to simplified models represented by $C_\alpha$ and
$C_\beta$ atoms, or by  backbone atoms only.

\renewcommand{\arraystretch}{.5}
\begin{table}[htb] % one column table
\caption[Performance of CALSP on Levitt's decoys]{\sf Performance of
CALSP for three decoy sets
\cite{ParkLevitt96_JMB,Levitt83_JMB,SamudralaLevitt99_PSB,XiaLevitt00_JCP}.
The numbers are the ranking of the native proteins. Results not
available from the references are labeled as ``N/A''.}
\label{tab:dec_pfm} \vspace*{.05in} \begin{center}
\begin{tabular}{|l|r|r|r|r|r|}
\hline Decoy sets  & CALSP     & LL[8]        & TE13     & MJ \\
\hline
A) 4state \cite{ParkLevitt96_JMB}   &  &  &  &  \\
1ctf    & 1 & 1 & 1 & 1 \\
1r69    & 1 & 1 & 1 & 1 \\
1sn3        & 2        & 1        &  6          &  2  \\
2cro        & 2             &  1        & 1     &  1  \\
3icb        & 1         & 5       &  N/A        & N/A  \\
4pti        & 2           & 1      &  7         & 3   \\
4rxn        & 3          & 51      & 16         & 1   \\ \hline
B) LMDS \cite{Levitt83_JMB}    &         &       &   &  \\
1b0n-B      & 1            & 2     & N/A         & N/A  \\
1bba        & 436          & 217    & N/A        & N/A  \\
1ctf & 1 & 1 & 1 & 1 \\
1fc2        & 83         & 500     & 14         & 501 \\
1dtk        & 1           & 2     & 5           & 13  \\
1igd        & 1          & 9       & 2          & 1   \\
1shf-A      & 3          & 17      & 1          & 11  \\
2cro & 1 & 1 & 1 & 1 \\
2ovo        & 4           & 3     & 1           & 2   \\
4pti        & 1          & 9     & N/A           & N/A  \\ \hline
C) lattice\_ssfit \cite{SamudralaLevitt99_PSB,XiaLevitt00_JCP} &  & & & \\
1beo        & 1           &  1     & N/A         & N/A  \\
1ctf & 1 & 1 & 1 & 1 \\
1dkt-A      & 1           &  1     & 2          & 32  \\
1fca        & 7            &  40     & 36       & 5   \\
1nkl & 1 & 1 & 1 & 1 \\
1trl-A      & 56           &  5    & 1          & 4   \\
1pgb & 1 & 1 & 1 & 1 \\
4icb        & 1          & 1      & N/A          & N/A  \\ \hline
 \end{tabular}
\end{center}
\end{table}

\section{Conclusion and Discussion}

{\bf Discrete state representation.}  In this study, we aim to develop
an effective potential function for simplified protein models. We use
discrete state model for representation of protein structures.  We
obtained discrete state values of bond angle $\alpha$ and torsion
angle $\tau$ from 3 to 10 states (see supplementary data). By
generating near native structures of low RMSD to native structures
using the discrete state model, we show that these models lead to
accurate modeling of native proteins.

The discrete state representation can provide a concise way to
represent protein structures by a sequence of states. Unlike
representation of secondary structure types (such as $H$ for helices,
$E$ for $\beta$ strand, and $C$ for coil and turns), the sequence of
discrete states at each residue position uniquely determines the three
dimensional conformation.  Methods in secondary structure prediction
are well developed with prediction accuracy as high as 80\%
\cite{PSIPRED}. Prediction of discrete states can benefit from
algorithms developed for secondary structure prediction
\cite{KuangYang04_Bioinfo}.  Predicted discrete state may be more
useful for tertiary structure prediction than predicted secondary
structures, since the residue specific discrete states are more
informative. Although there have been various attempts to define
secondary structure types other than the three basic types, we argue
that discrete states provide a natural and flexible representation,
where a different number of discrete states can be used for different
amino acids.  This provides a wide range of models with different
complexity and accuracy for studying proteins.

{\bf Reduced amino acid residue alphabet.}  We have simplified
amino acid residue alphabet using neighboring residue interactions
measured by the first order state transition probability. Hierarchical
clustering divides all amino acids into two groups with PRO (proline)
and GLY (glycine) separated from the rest of amino acids. Clearly,
geometric properties rather than chemical properties dominates at
the top level of the clustering.  Glycine is very small and
the distribution of its ($\alpha$, $\tau$) angles has more accessible
regions than any other residue types. On the other hand, proline has a
very rigid side chain and the distribution of its ($\alpha$, $\tau$)
angles has very limited accessible regions than other amino acid
types. These two amino acids are indeed found more frequently in
turn/loop conformations among these three secondary structure types
($\alpha$, $\beta$, and turns) \cite{CrastoFeng01_P,XiaXie02_MBE}. The
remaining amino acids are clustered into two groups with one group
being \{ILVYFWCM\} and the other group being \{DNHSTEKAQR\}, which
follows roughly their hydrophobicity at the second level of
clustering. This indicates that the local neighboring interaction is
also affected significantly by the hydrophobicity of the amino acids
\cite{AvbeljBaldwin04_PNAS}.

Geometric properties also play important roles in the detailed
clustering of amino acids both in the hydrophobic group and the polar
group of amino acids. For example, N (Asparagine) and D (Aspartic
acid) are clustered together and have large distances to the other
polar amino acids.  This is quite different from other clustering
results based on chemical properties or mutational propensities.  This
difference is probably due to the fact that both of these amino acids
can form favorable intra-residue hydrogen bond between their main
chains and the polar group on the side chain.  This would affect
significantly the geometry of their backbones.  Compared with two
other similar amino acids, E (Glutamic acid) and Q (Glutamine), D
and N are preferred for turn/loop conformations
\cite{CrastoFeng01_P,XiaXie02_MBE}. The simplification of amino acids
based on local sequence-structure propensities observed in native
proteins provides an alternative simplified amino acid alphabet, which
will be useful for representation and geometric modeling of protein
structures. This simplified alphabet would also be useful in building
fragment libraries and in predicting local structures from sequences.

{\bf Choice of descriptors.}  The choice of a specific set of
descriptor is critical for the success of potential functions for
simplified representations.  With a fixed representation, it is always
desirable to extract as much useful information as possible by
choosing an appropriate set of descriptors.  Conversely, since the
extractable information from a particular structure are limited by the
representation, the consideration of descriptors always affect the
choice of representation.  Many new descriptors incorporating a
variety of different types of information have been developed, such as
atomic pair-wise contact calculated by alpha-shape method
\cite{LiHuLiang03_P} or Voronoi tessellation
\cite{McConkeySobolev03_PNAS}, distance dependent contact instead of
simple distance cut-off
\cite{BaharJernigan97_JMB,TobiElber00_P,LuSkolnick01_P}, contact order
dependent contact \cite{Betancourt03_P}, and secondary structure
dependent contact \cite{SimonsBaker99_P}. We replaced the 210 contact
descriptors by distance-dependent contact descriptors with several
different distance intervals, but did not observe any noticeable
improvement. This is probably because of the limitation of the contact
information that can be extracted from simplified representation.

{\bf Improving potential function for a fixed representation.}
Potential functions using weighted linear combination of residue-level
contact descriptors defined by simple distance cut-offs have been
shown to be inadequate in discriminating many native structures from a
large number of decoys
\cite{VendruscoloDomany00_P,HuLiLiang04_Bioinfo}. Among many possible
improvements, the modification of descriptors without changing the
representation is convenient.  Recent study on the conformational
biases used in Monte Carlo simulations by several successful folding
methods suggest that such conformational biases likely serve as an
energy term missing in current potential function
\cite{Przytycka04_P}.  Our work can be regarded as an effort in
searching for the missing information of the potential function. In
this study, we combine local sequence-structure descriptors with
contact descriptors, while keeping both the functional form and
representation as simple as possible. Although the current local
sequence-structure descriptors are quite simple, the performance of
the potential function has been significantly improved compared to the
one using only contact propensity.

It is likely that the best potential function will be different if
a different protein model is used.  Our potential function can be
adapted for use with other protein models generated by different
sampling methods.  The discrimination surface between native
proteins and decoys is determined by points (namely, proteins and
decoys) along the boundary surface \cite{HuLiLiang04_Bioinfo}.
This surface is determined by the protein model and the method of
structure generation, but is invariant once the model and the
method are fixed. Therefore, it is necessary to develop different
optimized potential functions for different protein models.  An
improved potential function can be obtained by adding new decoys
that are challenging for this particular protein model to the
training set. This case-by-case approach is also practical. In
applications where a potential function is used to discriminate
native structures from decoy structures, a researcher usually
decides upon choosing a favorite protein model and a
structure-generation method, as is the case in research works of
protein structure prediction.  Since the user has access to a
method to generate a large number of candidate structures, decoy
structures can be easily obtained for training an improved
potential function. This improved potential function can be based
on the descriptors and functional form of the original CALSP
potential.  The effectiveness of this approach can be seen from
the performance of our potential functions on LKF and Baker decoy
sets, where only decoy structures from LKF and Baker decoy sets
are used in training.

Although it is impossible to develop a one-size-fits-all potential
function for all simplified models, our study showed that the new set
of descriptors, the method for their simplification, and the simple
functional form to combine them are generally applicable to other
protein models. Our study suggests a novel approach to develop
effective potential functions for simplified protein models.

{\bf Further improvement of potential function.}  Potential
functions RCALSP1 (using reduced amino acid alphabet in deriving
contact descriptors) and RCALSP2 (further incorporating sequence
separation information in the descriptors) show slightly better
performance compared to CALSP in discrimination of gapless
threading decoys, even though they have reduced numbers of
parameters, {\it i.e.}, 455 for RCALSP1 and 565 for RCALSP2,
compared to 610 for CALSP. Although more detailed studies are
needed to assess the effectiveness of these two potential
functions, these results point to a promising direction to further
improve the potential function.  We expect that many local
sequence-structure descriptors are redundant, {\it e.g.}, some
local sequences have no preference for local structure. This
indicates that the current set of descriptors can be further
simplified, which will provide additional rooms for incorporation
of more informative descriptors. Identification of important
descriptors will also shed light on the determinants of protein
folding and stability.

The local sequence-structure propensity currently used considers
only two adjacent residues on the sequence, which cannot capture
more complex local interactions beyond the two neighboring
residues. Additional descriptors can be derived from two residues
not adjacent on the sequence or from more than two residues. The
addition of more descriptors will need to be done carefully to
avoid the over-fitting problem.

{\bf Local sequence-structure relationship and protein folding.}
Experimental study has shown that the unfolded state of proteins still
maintain much of the native topology under strong denaturing condition
\cite{Shortle01_S}. The origin of interactions between neighboring
residue has been studied recently by electrostatic calculations of
peptide solvation \cite{AvbeljBaldwin04_PNAS}. Our results in
clustering amino acids based on their sequence neighbor interactions
suggests that this mainly originated from the geometric properties of
amino acids, but is also significantly influenced by their
physicochemical properties.  Experimental studies and successful
application of local sequence-structure relationship in structure
predictions clearly indicate that local sequences or sequence
fragments also have strong preference for adopting certain native
local structure. Such local sequence-structure relationship could be
important for decreasing the large entropy during the folding process.
Local sequence-structure correlations induced by neighboring residue
interactions may play important roles in the unfolded states, such
that the majority of unfolded conformation may be located around the
native conformation, although not in the sense of close RMSD
\cite{FitzkeeRose04_PNAS}.  The realization of local and strong
sequence-structure correlations may induce more distant but weaker
sequence-structure correlations spontaneously at many locations on the
peptide sequence, which could be a part of the cooperative folding
process.  Therefore, the entropy of unfolded state may be dramatically
reduced at the very beginning of protein folding due to the
correlations between local sequence and structure, which can take
effect even during the protein synthesis. The folding entropy,
considered as the major force opposing protein folding, therefore may
not be so large as thought before.

{\bf Summary.} Simplified protein representation can effectively
reduce the conformational search space and provides an attractive
model for computational studies of protein structures.  However,
currently there are no empirical potential functions that are
applicable for simplified protein model.  In this work, we develop
empirical potential function for simplified protein models by
combining descriptors derived from residue-residue contact and
local sequence-structure relationship.  The parameters are
obtained by optimizing discrimination of native proteins and
decoys.  Based on testing with a variety of decoy sets, our
results show that this strategy is effective, and the empirical
potentials developed here requiring only $C_\alpha$ or backbone
atoms have better or similar performance in decoy discrimination
compared to other residue-level potentials requiring in addition
either full atom structure or models of side chains.  We also
showed that for a large representative set of proteins, discrete
state model can be very accurate.  We also found that the
conformations of nearest sequence neighbors often strongly
influence each other, and such correlation can be employed to
provide additional discrimination in addition to contact
interactions.  We further develop reduced alphabet of amino acids
based on analysis of local sequence-structure correlation of
neighboring residues.  The results indicate that there are
characteristic properties in adopting local conformations among
groups of residues, and such grouping is different from grouping
based on contact interactions. We showed that reduced alphabet
helps to improve discrimination.  The rich information contained
in local sequence-structure descriptors suggest that local effects
may play important role in reducing entropic cost in protein
folding.

{\bf Note.}  Details of the parameters of the potential functions, the
angles for the reduced state models of amino acid residues, and
a list of the set of 978 proteins can all be found at ({\tt
gila.bioengr.uic.edu/pub-data/potential05-proteins/}).

\section{Acknowledgment}
This work is supported by grants from NSF (CAREER DBI0133856),
NIH (GM68958), and ONR (N000140310329).

\end{document}